\documentstyle[emulateapj,epsf]{article}

\lefthead{Ogilvie \& Lubow}
\righthead{Three-dimensional waves in an accretion disk}

\begin{document}

\title{The effect of an isothermal atmosphere on the propagation of
three-dimensional waves in a thermally stratified accretion disk}

\author{G.~I.~Ogilvie\altaffilmark{1,3} and S.~H.~Lubow\altaffilmark{2,3}}
\altaffiltext{1}{Institute of Astronomy, University of Cambridge, Madingley
Road, Cambridge CB3~0HA, United Kingdom}
\altaffiltext{2}{Space Telescope Science Institute, 3700 San Martin Drive,
Baltimore, MD~21218}
\altaffiltext{3}{Isaac Newton Institute, University of Cambridge, 20
Clarkson Road, Cambridge CB3~0EH, United Kingdom}

\slugcomment{Submitted to the Astrophysical Journal}

\begin{abstract}
We extend our analysis of the three-dimensional response of a
vertically polytropic disk to tidal forcing at Lindblad resonances by
including the effects of a disk atmosphere.  The atmosphere is modeled
as an isothermal layer that joins smoothly on to an underlying
polytropic layer.  The launched wave progressively enters the
atmosphere as it propagates away from the resonance.  The wave never
propagates vertically, however, and the wave energy rises to a
(finite) characteristic height in the atmosphere.  The increase of
wave amplitude associated with this process of wave channeling is
reduced by the effect of the atmosphere.  For waves of large azimuthal
mode number $m$ generated by giant planets embedded in a disk, the
increase in wave amplitude is still substantial enough to be likely to
dissipate the wave energy by shocks for even modest optical depths
($\tau\ga10$) over a radial distance of a few times the disk
thickness.  For low-$m$ waves generated in circumstellar disks in
binary stars, the effects of wave channeling are less important and
the level of wave nonlinearity increases by less than a factor of 10
in going from the disk edge to the disk center.  For circumbinary
disks, the effects of wave channeling remain important, even for
modest values of optical depth.

\end{abstract}


\section{Introduction}

Tidally generated waves can play an important role in the dynamics of
gaseous disks and the tidal perturbers.  The Lindblad resonances (LRs)
are important sites of wave generation in disks (Goldreich \& Tremaine
1979).  Most previous studies have approximated the disk as
two-dimensional (2D) by ignoring effects in the vertical direction
(perpendicular to the orbit plane).  For disks whose structure and
thermodynamic response are locally vertically isothermal, the 2D
approximation accurately describes the wave that is generated at an
LR, which is indeed a 2D sound wave.  In many cases of astrophysical
interest, however, the disk is thermally stratified (optically thick)
in the vertical direction.  Such cases include the planet-forming
regions of protostellar disks (Bell et~al. 1997) and cataclysmic
variable (CV) disks (La Dous 1994).

In an earlier paper (Lubow \& Ogilvie 1998, hereafter Paper~I), we
analyzed the linear response of a thin, vertically polytropic disk to
tidal forcing at LRs.  We found that the dominant mode of excitation
was the f mode of even symmetry.  Close to the resonance, this mode
behaves compressibly and occupies the full vertical extent of the
disk, like the 2D mode in an isothermal disk.  However, as the wave
propagates away from the resonance through a radial distance of order
$r_{\rm L}/m$ (for resonance radius $r_{\rm L}$ and azimuthal mode
number $m$), it behaves incompressibly, like a surface gravity mode,
and becomes confined close to the disk surface.  The degree of
confinement increases with distance from the resonance.  As a
consequence of this wave-channeling process, there is usually a strong
increase in wave amplitude with radial distance from the LR on the
scale $r_{\rm L}/m$.  Although this analysis did not include damping
or nonlinear effects explicitly, we estimated that shocks would
sometimes develop and damp the wave.  Other damping mechanisms such as
radiative damping (Cassen \& Woolum 1996) or viscous damping may also
be of importance.

Once the wave energy is channeled to a region close to the disk
surface, the wave's properties are determined by the structure of the
disk's outer layers.  In particular, the effects of a disk atmosphere
can be important. Such effects are not well represented by a
vertically polytropic disk.  In the case of a polytrope, there is a
definite surface on which the density and temperature drop to zero.
With an atmosphere, the disk density drops more gradually with height
and the temperature is approximately constant. In fact, there is no
actual disk surface in the isothermal case.  The density falls as a
Gaussian with height above the mid-plane.  In the limit of very high
(vertical) optical depth in the disk, the polytropic approximation
should be adequate because the atmosphere contains an insignificant
fraction of the disk's mass.  In the opposite limit of an optically
thin disk, the disk structure is vertically isothermal and the wave
launched at resonance behaves as a 2D sound wave.

The purpose of this paper is to determine to consequences of a finite,
non-zero optical depth on the propagation of a wave launched from an
LR. For disks of optical depth of about 100, which is thought to be
typical in planet-forming regions of protostellar disks, nearly one
per cent of the disk's mass resides in the atmosphere.  As a result of
wave channeling, the wave energy could be contained within the
atmosphere.  The purpose of this paper is to explore the consequences
of a disk atmosphere on the propagational properties of the waves
generated at LRs.  We aim to determine whether the wave in the disk
atmosphere would propagate upwards into very low density regions where
shock dissipation would occur.  More generally, we want to determine
how much the wave amplitudes can be amplified by the wave-channeling
process in the presence of a disk atmosphere as a function of disk
optical depth.

\section{Description of the model}

The semi-analytic approach taken in Paper~I can be extended to include
the disk atmosphere, which is modeled as an isothermal layer that
extends from some height above the mid-plane outward. Beneath this
layer resides a thermally stratified polytropic layer which extends
vertically down to the mid-plane.  For convenience, we treat the
transition as abrupt, although in reality it may occur over a distance
comparable to a scale height.

As will be shown, physical solutions to the wave equation in the
isothermal layer can be obtained analytically.  The upper boundary
condition of the isothermal layer is that no waves carry energy from
great distances above the disk mid-plane downwards toward the
mid-plane. That is, there are no external sources of energy above the
disk.  The solutions allow for the possibility of reflection in the
disk atmosphere and propagation to great (or infinite) heights in the
atmosphere.

These atmospheric solutions contain some freedom that needs to be
constrained by the underlying polytropic layer.  The isothermal
solutions provide constraints that serve as an outer boundary
condition for the numerical solution of the wave equations in the
polytropic layer.  Taken together, vertically global solutions are
obtained that connect the two layers in a physically meaningful
manner. The ratio of the temperature at mid-plane to the temperature
at the layer interface determines the disk optical depth through the
$T$--$\tau$ relation.

\section{Equilibrium of the disk}

The disk is considered to be a steady, axisymmetric fluid rotating in
a constant, axisymmetric gravitational potential.  Referred to
cylindrical polar coordinates $(r,\phi,z)$, the fluid has angular
velocity $\Omega(r,z)$, density $\rho(r,z)$, and pressure $p(r,z)$,
while the gravitational potential is $\Phi(r,z)$.  The disk is
considered to be thin, so that the angular velocity can be considered
independent of $z$ and given by
\begin{equation}
r\Omega^2={{\partial\Phi}\over{\partial r}}\bigg|_{z=0},
\end{equation}
while the epicyclic frequency $\kappa$ is given by
\begin{equation}
\kappa^2=4\Omega^2+2r\Omega{{\rm d}\Omega\over{\rm d}r}.
\end{equation}
The effective gravitational acceleration is $-g(r,z)\,{\bf e}_z$, where
\begin{equation}
g=\Omega_\perp^2z,
\end{equation}
and $\Omega_\perp(r)$ is the vertical frequency given by
\begin{equation}
\Omega_\perp^2={\partial^2\Phi\over\partial z^2}\bigg|_{z=0}.
\end{equation}
In particular, we will be interested in a Keplerian disk about a
spherical mass $M$, for which the potential is
\begin{equation}
\Phi=-GM(r^2+z^2)^{-1/2}
\end{equation}
and $\kappa=\Omega_\perp=\Omega$.  However, the notational distinction
between $\Omega$, $\kappa$, and $\Omega_\perp$ is retained, partly to
aid the interpretation of the equations in the case of a Keplerian
disk, and partly to allow a generalization to non-Keplerian disks.

The local vertical equilibrium of the disk is determined by the
equation
\begin{equation}
{{\partial p}\over{\partial z}}=-\rho g.
\end{equation}
The buoyancy frequency $N(r,z)$ of vertical oscillations is given by
\begin{equation}
N^2=g\left({{1}\over{\gamma p}}{{\partial p}\over{\partial
z}}-{{1}\over{\rho}}{{\partial\rho}\over{\partial z}}\right),
\end{equation}
where $\gamma$ is the adiabatic exponent.  Throughout this paper, the
self-gravitation of the fluid, its viscosity, and the accretion flow
are neglected.

In order to model a disk of finite optical depth, we assume that the
pressure and density are related by a polytropic relation,
\begin{equation}
p=K\rho^{1+1/s},
\end{equation}
in a central layer $|z|<z_1$ containing the mid-plane, and by an
isothermal relation,
\begin{equation}
p=c^2\rho,
\end{equation}
in the outer layers $|z|>z_1$.  The pressure and density are required
to be continuous at the interface.  The solution is then
\begin{eqnarray}
\hskip-0.5cm|z|<z_1:&&\hskip-0.5cm\rho=\rho_0\left(1-{{z^2}\over{H_s^2}}
\right)^s,\\
\hskip-0.5cm|z|>z_1:&&\hskip-0.5cm\rho=\rho_0\left(1-{{z_1^2}\over{H_s^2}}
\right)^s
\exp\left[-{{(z^2-z_1^2)}\over{2H^2}}\right]
\end{eqnarray}
for the density, and
\begin{eqnarray}
\hskip-0.5cm|z|<z_1:&&\hskip-0.5cm p=p_0\left(1-{{z^2}\over{H_s^2}}
\right)^{s+1},\\
\hskip-0.5cm|z|>z_1:&&\hskip-0.5cm p=p_0\left(1-{{z_1^2}\over{H_s^2}}
\right)^{s+1}
\exp\left[-{{(z^2-z_1^2)}\over{2H^2}}\right]
\end{eqnarray}
for the pressure.  Here $\rho_0(r)$ and $p_0(r)$ are the density and
pressure on the mid-plane, and $H=c/\Omega_\perp$ is the isothermal
scale height.  The quantity $H_s$ is the height at which the
polytropic layer would reach zero density and pressure if it were not
truncated, and is given by
\begin{equation}
H_s^2={{2(s+1)p_0}\over{\Omega_\perp^2\rho_0}}.
\end{equation}
Furthermore, the relation
\begin{equation}
H_s^2-z_1^2=2(s+1)H^2
\end{equation}
holds, because $dp/dz$ is continuous at the interface.

The quantity $z_1/H_s$ is a measure of the `weighting' assigned to the
polytropic layer in this mixed model.  The ratio of the temperature of
the isothermal layer to the temperature at the mid-plane (for an ideal
gas) is
\begin{equation}
{T_{\rm i}\over{T_{\rm m}}}=1-{{z_1^2}\over{H_s^2}}.
\end{equation}
This may be expressed in terms of an approximate effective optical
depth $\tau$ at the mid-plane according to the relation (e.g.~Bell
et~al. 1997)
\begin{equation}
T_{\rm m}^4={\textstyle{{3}\over{8}}}\tau T_{\rm
i}^4,
\end{equation}
which gives
\begin{equation}
\tau={\textstyle{{8}\over{3}}}
\left(1-{{z_1^2}\over{H_s^2}}\right)^{-4}.
\end{equation}

The ratio $\Sigma_{\rm i}/\Sigma_{\rm p}$ of the surface densities of
the two layers can be expressed in terms of $z_1/H_s$, but only by
using higher transcendental functions.

We assume that the effective adiabatic exponent $\gamma$ is constant
and greater than unity in the central layer, but equal to unity in the
outer layers.  This reflects the fact that the thermal time-scale is
short compared to the dynamical time-scale in the isothermal layers.

The buoyancy frequency may then be evaluated as
\begin{eqnarray}
\hskip-0.5cm|z|<z_1:&&\hskip-0.5cm N^2=\left[s-\left({{s+1}\over{\gamma}}
\right)\right]
\left({{2z^2}\over{H_s^2-z^2}}\right)\Omega_\perp^2,\\
\hskip-0.5cm|z|>z_1:&&\hskip-0.5cm N^2=0,
\end{eqnarray}
and is discontinuous at $|z|=z_1$.

\section{Description of the local dispersion relation}

\subsection{Basic equations}

The equations governing free, linear waves in a thin accretion disk
have been derived and solved in different cases by Lubow \& Pringle
(1993, hereafter LP), Korycansky \& Pringle (1995), Ogilvie (1998),
and in Paper~I.  The separation of scales between the horizontal and
vertical directions allows all wave quantities to be expressed as WKB
functions in $r$, except in the neighborhood of resonances, which are
also turning points for the waves.  In particular, a wave quantity $X$
assumes the form
\begin{eqnarray}
\lefteqn{\hskip-1cm X(r,\phi,z,t)\sim{\rm Re}\left\{\tilde
X(r,z)\right.}&\nonumber\\
&&\left.\hskip-1cm\times\exp\left[-i\omega
t+im\phi+i\int k(r)\,dr\right]\right\},
\end{eqnarray}
where $\omega$ is the frequency eigenvalue, $m$ is the azimuthal
wavenumber and $k(r)$ is the radial wavenumber.  The tilde will be
omitted hereafter.  The equations satisfied by the Eulerian velocity
perturbation $(u,v,w)$ and pressure perturbation $p^\prime$ at each
radius may be written
\begin{equation}
\rho(\hat\omega^2-\kappa^2)u=\hat\omega kp^\prime,\label{eq1}
\end{equation}
\begin{equation}
\rho(\hat\omega^2-N^2)w=-i\hat\omega\left({{\partial p^\prime}\over{\partial
z}}+{{gp^\prime}\over{v_{\rm s}^2}}\right),
\end{equation}
and
\begin{equation}
-i\hat\omega p^\prime=-\gamma p\left(iku+{{\partial w}\over{\partial
z}}\right)+\rho gw,\label{eq2}
\end{equation}
where $v_{\rm s}=(\gamma p/\rho)^{1/2}$ is the sound speed and
\begin{equation}
\hat\omega=\omega-m\Omega
\end{equation}
is the intrinsic frequency of the wave.  The azimuthal velocity
perturbation is related to the radial one by
\begin{equation}
-i\hat\omega v+2Bu=0,
\end{equation}
where
\begin{equation}
B=\Omega+{{r}\over{2}}{{d\Omega}\over{dr}}
\end{equation}
is the usual Oort parameter.

\subsection{Analytical solutions in the isothermal layer}

LP solved analytically for the vertical structure of modes in a purely
isothermal disk by obtaining series solutions of the equations about
$z=0$.  The relevant solutions were identified as terminating power
series (i.e. polynomials) of either even or odd symmetry, multiplied
by an exponential factor.  In the present case, we must consider more
general solutions because different boundary conditions apply at the
interface between the two layers.

In the isothermal layer, equations (\ref{eq1})--(\ref{eq2}) may be
rewritten as
\begin{equation}
{{\partial u}\over{\partial
z}}=\left({{\hat\omega^2}\over{\hat\omega^2-\kappa^2}}\right)
ikw\label{eq3}
\end{equation}
and
\begin{equation}
{{\partial w}\over{\partial
z}}=\left({{1}\over{ik}}\right)\left[k^2
-\left({{\hat\omega^2-\kappa^2}\over{\Omega_\perp^2}}\right)
{{1}\over{H^2}}\right]u+\left({{z}\over{H^2}}\right)w.\label{eq4}
\end{equation}
There exists a 2D mode consisting of a purely horizontal
motion independent of $z$, which satisfies the dispersion relation
\begin{equation}
\hat\omega^2=\kappa^2+c^2k^2.
\end{equation}
For all other modes, equations (\ref{eq3}) and (\ref{eq4}) can be
combined to give a single, second-order equation for $w$ in the form
\begin{eqnarray}
\lefteqn{\hskip-0.5cm{{\partial^2w}\over{\partial
z^2}}-\left({{z}\over{H^2}}\right){{\partial w}\over{\partial
z}}}&\nonumber\\
&&\hskip-0.5cm+\left[\left({{\hat\omega^2-\Omega_\perp^2}
\over{\Omega_\perp^2}}\right){{1}\over{
H^2}}-\left({{\hat\omega^2}\over{\hat\omega^2-\kappa^2}}\right)k^2
\right]w=0.\label{eq5}
\end{eqnarray}
This equation can be transformed into the parabolic cylinder equation
(Abramowitz \& Stegun 1965, hereafter AS).  We will be concerned with
solutions for which both $k$ and $\hat\omega$ are real.

\subsubsection{Transformation of the equation}

The substitution
\begin{equation}
w(z)=y(x)\exp\left({\textstyle{{1}\over{4}}}x^2\right),
\end{equation}
where
\begin{equation}
x={{z}\over{H}},
\end{equation}
transforms equation (\ref{eq5}) into
\begin{equation}
{{d^2y}\over{dx^2}}-\left({\textstyle{{1}\over{4}}}x^2+a\right)y=0,
\end{equation}
where
\begin{equation}
a=\left({{\hat\omega^2}\over{\hat\omega^2-\kappa^2}}\right)
k^2H^2-\left({{\hat\omega^2-\Omega_\perp^2}
\over{\Omega_\perp^2}}\right)-{\textstyle{{1}\over{2}}}.
\end{equation}
Two linearly independent solutions are $U(a,x)$ and $V(a,x)$, defined
by AS.  The asymptotic forms as $x\to+\infty$ are (AS, eq. [19.8])
\begin{eqnarray}
U(a,x)&\sim&x^{-(a+1/2)}\exp\left(-{\textstyle{{1}\over{4}}}x^2\right),\\
V(a,x)&\sim&\left({{2}\over{\pi}}\right)^{1/2}x^{a-1/2}
\exp\left({\textstyle{{1}\over{4}}}x^2\right).
\end{eqnarray}

\subsubsection{Conditions at $z=\pm\infty$}

In order to select physically acceptable solutions, we must examine
the behavior of the wave action as $|z|\to\infty$.  A suitable energy
wave action can be defined, whose density, averaged over time, is
\begin{equation}
{\cal A}^{\rm(e)}={{\rho\omega}\over{2\hat\omega}}
\left(|u|^2+|w|^2\right),
\end{equation}
and whose vertical flux, also averaged over time, is
\begin{equation}
{\cal F}^{\rm(e)}_z={\rm Re}\left(w^*p^\prime\right).
\end{equation}

In the case of the isothermal layer, the limiting behavior of the wave
action as $z\to+\infty$ is
\begin{eqnarray}
{\cal A}^{\rm(e)}&\propto&x^{1-2a}
\exp\left(-{\textstyle{{1}\over{2}}}x^2\right),\\
{\cal
F}^{\rm(e)}_z&\propto&x^{-2a}\exp\left(-{\textstyle{{1}\over{2}}}x^2\right)
\end{eqnarray}
for the solution $U(a,x)$, and
\begin{eqnarray}
{\cal A}^{\rm(e)}&\propto&x^{1+2a}
\exp\left({\textstyle{{1}\over{2}}}x^2\right),\\
{\cal
F}^{\rm(e)}_z&\propto&x^{2a}
\exp\left({\textstyle{{1}\over{2}}}x^2\right)
\end{eqnarray}
for the solution $V(a,x)$.  It is clear that only the solution
$U(a,x)$ is acceptable at $z=+\infty$.  Similarly, only the solution
$U(a,-x)$ is acceptable at $z=-\infty$.  These solutions represent
evanescent waves at infinity.

\subsubsection{Modes in a purely isothermal disk}

When the disk is purely isothermal without a polytropic layer, we
recover the results of LP as follows.  An acceptable solution for
$y(x)$ must be proportional to both $U(a,x)$ and $U(a,-x)$.  However,
these functions are linearly dependent only when
$a=-n-{\textstyle{{1}\over{2}}}$, with $n$ a non-negative integer.  In
that case
\begin{equation}
U(-n-{\textstyle{{1}\over{2}}},x)=2^{-n/2}H_n\left(2^{-1/2}x\right)
\exp\left(-{\textstyle{{1}\over{4}}}x^2\right),
\end{equation}
where $H_n$ is the Hermite polynomial of degree $n$ (AS,
eq. [19.13.1]).  This condition leads to the dispersion relation
\begin{equation}
n=\left({{\hat\omega^2-\Omega_\perp^2}\over{\Omega_\perp^2}}\right)
-\left({{\hat\omega^2}\over{\hat\omega^2-\kappa^2}}\right)k^2H^2\label{eq6}
\end{equation}
equivalent to equation (54) of LP for the case $\gamma=1$.

\subsection{Numerical solutions in the polytropic layer}

The equations in the polytropic layer must be solved numerically, as
in Korycansky \& Pringle (1995).  In a form similar to equations
(\ref{eq3}) and (\ref{eq4}), they are
\begin{eqnarray}
\lefteqn{\hskip-0.5cm{{\partial u}\over{\partial
z}}=\left[s-\left({{s+1}\over{\gamma}}\right)\right]\left({{2z}
\over{H_s^2-z^2}}\right)u+\left({{ik}\over{\hat\omega^2-\kappa^2}}\right)}
&\nonumber\\
&&\hskip-0.5cm\times\left\{\hat\omega^2-\Omega_\perp^2\left[s-\left({{s+1}
\over{\gamma}}\right)
\right]\left({{2z^2}\over{H_s^2-z^2}}\right)\right\}w\label{eq9}
\end{eqnarray}
and
\begin{eqnarray}
\lefteqn{\hskip-2cm{{\partial w}\over{\partial z}}=\left({{1}\over{ik}}
\right)\left[k^2-\left
({{s+1}\over{\gamma}}\right)\left({{\hat\omega^2-\kappa^2}\over
{\Omega_\perp^2}}\right)\left({{2}\over{H_s^2-z^2}}\right)\right]u}&
\nonumber\\
&&+\left({{s+1}\over{\gamma}}\right)\left({{2z}\over{H_s^2-z^2}}\right)w.
\label{eq10}
\end{eqnarray}
The boundary conditions at $z=0$ are the usual symmetry
conditions,
\begin{equation}
{{\partial u}\over{\partial z}}=w=0
\end{equation}
for an even mode, and
\begin{equation}
u={{\partial w}\over{\partial z}}=0
\end{equation}
for an odd mode.  At $z=z_1$, the solutions must be matched on to the
solutions in the isothermal layer.  Both $u$ and $w$ must be
continuous.

The matching condition is
\begin{equation}
{{u}\over{w}}=ikH\left[k^2H^2-
\left({{\hat\omega^2-\kappa^2}\over{\Omega_\perp^2}}\right)\right]^{-1}
\left[{{U^\prime(a,x_1)}\over{U(a,x_1)}}-{{x_1}\over{2}}\right].\label{eq8}
\end{equation}
where $x_1=z_1/H$, and the prime denotes differentiation with respect
to the (second) argument.  (If this ratio diverges, the matching
condition is $w=0$.)

The numerical method is as follows.  A value of $k$ and a symmetry
(either even or odd) are chosen.  The values of $u$ and $w$ at $z=0$
are determined by the appropriate symmetry condition and by an
arbitrary normalization condition.  The value of $\hat\omega$ must be
guessed.  The equations are then integrated from $z=0$ to $z=z_1$, and
the value of $\hat\omega$ tuned so that the matching condition is
satisfied.

To evaluate the function $U(a,x)$ and its derivative numerically, we
use the following method.  When $a>0$ and $x^2+4a\gg1$, Darwin's
expansion (AS, eq. [19.10.2]) provides an accurate asymptotic
approximation.  If these conditions are not satisfied, we choose a
positive integer $N$ such that $U(a+N,x)$ and $U(a+N+1,x)$ can be
evaluated accurately using Darwin's expansion.  These values are used
to initialize the recurrence relation
\begin{equation}
U(a+n-1,x)=xU(a+n,x)+(a+n+{\textstyle{{1}\over{2}}})U(a+n+1,x)
\end{equation}
(AS, eq. [19.6.4]), which is then iterated to determine $U(a,x)$.
(The recurrence is stable in this direction.)  Finally, the recurrence
relation
\begin{equation}
U^\prime(a,x)=-{\textstyle{{1}\over{2}}}xU(a,x)-
(a+{\textstyle{{1}\over{2}}})U(a+1,x)
\end{equation}
(AS, eq. [19.6.1]) provides the derivative.

\subsection{Asymptotic solutions in the limit $kH\to\infty$}

As waves propagate radially away from the resonances where they are
excited, the dispersion relation is often followed into a limit in
which $kH$ is large.  This can lead to wave channeling, as described
in Paper~I, and it is important to determine the behavior of the modes
in this limit so that the effects of nonlinear dissipation can be
estimated.

In a purely polytropic disk, Ogilvie (1998) showed that the f, p, and
g modes all become trapped near the surfaces of the disk in this
limit, and have $\hat\omega^2/\Omega_\perp^2=O(kH)$.  In contrast, the
r modes become trapped near the mid-plane, and have
$\hat\omega^2/\Omega_\perp^2=O\left((kH)^{-1}\right)$.  (In the
exceptional case when the disk is marginally stable to convection, the
r modes have $\hat\omega^2/\Omega_\perp^2=O\left((kH)^{-2}\right)$ and
do not become localized.)

In a purely isothermal disk, the limiting behavior of the dispersion
relation (\ref{eq6}) is easily found to be
\begin{equation}
\hat\omega^2=
c^2k^2+\kappa^2+(n+1)\Omega_\perp^2+O\left((kH)^{-2}\right)
\label{eq7}
\end{equation}
for the p modes, and
\begin{equation}
\hat\omega^2=(n+1){{\kappa^2}\over{k^2H^2}}+O\left((kH)^{-4}\right)
\end{equation}
for the r modes.  Evidently the 2D mode may be considered to
correspond to the case $n=-1$ in equation (\ref{eq7}).

For the mixed model considered in this paper, it is clear that the
limiting behavior of the r modes will agree with the purely polytropic
case and will not be affected by the outer layers (except in the
marginally stable case).  This cannot be true of the other modes,
however, since the surface at which they would have become
concentrated is no longer present.  Instead, the wave action moves out
into the isothermal layer.  We seek a solution in which, in accordance
with the behavior of all modes other than r modes in the purely
isothermal disk,
\begin{equation}
\hat\omega^2=c^2k^2+\kappa^2+({\textstyle{{1}\over{2}}}-\hat a)
\Omega_\perp^2+
O\left((kH)^{-2/3}\right),\label{eq12}
\end{equation}
where $\hat a$ is a constant to be determined.  (The scaling of the
remainder term will be seen later to be justified.)  Then
\begin{equation}
a=\hat a+O\left((kH)^{-2/3}\right).
\end{equation}
The matching condition at $z=z_1$ is, from equation (\ref{eq8}),
\begin{equation}
{{u}\over{w}}=-{{ikH}\over{({\textstyle{{1}\over{2}}}-\hat a)}}
\left[{{U^\prime(\hat a,x_1)}\over{U(\hat a,x_1)}}
-{{x_1}\over{2}}\right]+O\left((kH)^{1/3}\right).\label{eq11}
\end{equation}
Now consider the limiting form of the equations in the polytropic
layer.  A first examination of equations (\ref{eq9}) and (\ref{eq10})
yields
\begin{equation}
{{\partial u}\over{\partial z}}\sim ikw
\end{equation}
and
\begin{equation}
{{\partial w}\over{\partial
z}}\sim-ik\left({{z_1^2-z^2}\over{H_s^2-z^2}}\right)u,
\end{equation}
which imply
\begin{equation}
{{\partial^2u}\over{\partial z^2}}\sim
k^2\left({{z_1^2-z^2}\over{H_s^2-z^2}}\right)u.
\end{equation}
The solution in $z<z_1$ is an evanescent WKB function, but there is a
turning point at $z=z_1$.  In the neighborhood of the interface, the
solution must be described using an Airy function.  For
$(z_1-z)/z_1=O\left((kH)^{-2/3}\right)$, one finds
\begin{eqnarray}
u&\sim&Ai\left[\left({{2z_1}\over{H_s^2-z_1^2}}\right)^{1/3}k^{2/3}
(z_1-z)\right],\\
w&\sim&ik^{-1/3}\left({{2z_1}\over{H_s^2-z_1^2}}\right)^{1/3}\nonumber\\
&&\times Ai^\prime
\left[\left({{2z_1}\over{H_s^2-z_1^2}}\right)^{1/3}k^{2/3}(z_1-z)\right],
\end{eqnarray}
subject to an arbitrary normalization, and so
\begin{equation}
{{u}\over{w}}\sim-ik^{1/3}\left({{2z_1}\over{H_s^2-z_1^2}}\right)^{-1/3}
\left[{{Ai(0)}\over{Ai^\prime(0)}}\right]=O\left((kH)^{1/3}\right)
\end{equation}
at the interface.  For consistency with equation (\ref{eq11}), we
require
\begin{equation}
{{U^\prime(\hat a,x_1)}\over{U(\hat a,x_1)}}-{{x_1}\over{2}}=0.\label{eq13}
\end{equation}
This transcendental equation for $\hat a$ has infinitely many roots,
each of which corresponds to the limiting form of one branch of the
dispersion relation, according to equation (\ref{eq12}).

\section{Propagation of the f mode}

\subsection{Numerical results}

As in Paper~I, we focus of the properties of the ${\rm f}^{\rm e}$
mode, since it is by far the dominant mode excited at an LR.  In
obtaining numerical solutions, we have considered a Keplerian disk
whose polytropic layer has index $s=3$ and adiabatic exponent
$\gamma=5/3$.  We have examined four different values of the effective
optical depth: $\tau=10$, 100, 1000 and 10000.  The parameters of the
four models are summarized in Table~1.  The quantities given are the
ratio $\Sigma_{\rm i}/\Sigma_{\rm p}$ of the surface densities of the
isothermal and polytropic layers, the ratio $T_{\rm i}/T_{\rm m}$ of
the temperature of the isothermal atmosphere to the temperature at the
mid-plane, the dimensionless height $z_1/H_s$ of the interface, the
dimensionless scale height $H/H_s$ of the atmosphere, the fraction $f$
of the torque exerted at an LR that is carried by the ${\rm f}^{\rm
e}$ mode, and the eigenvalue $\hat a$ of equation (\ref{eq13})
corresponding to the ${\rm f}^{\rm e}$ mode.

\begin{deluxetable}{rrrrrrr}
\tablecaption{Parameters of the four models.\label{tab1}}
\tablewidth{0pt}
\tablehead{\colhead{$\tau$}&\colhead{$\Sigma_{\rm i}/\Sigma_{\rm p}$}
&\colhead{$T_{\rm i}/T_{\rm m}$}&\colhead{$z_1/H_s$}&\colhead{$H/H_s$}
&\colhead{$f$}&\colhead{$\hat a$}}
\startdata
$10$&$1.266\times10^{-1}$&$0.7186$&$0.5305$&$0.2997$&$0.9771$&$-2.558$\nl
$100$&$8.872\times10^{-3}$&$0.4041$&$0.7719$&$0.2248$&$0.9687$&$-5.751$\nl
$1000$&$8.025\times10^{-4}$&$0.2272$&$0.8791$&$0.1685$&$0.9679$&$-10.587$\nl
$10000$&$7.670\times10^{-5}$&$0.1278$&$0.9339$&$0.1264$&$0.9678$
&$-18.529$\nl
\enddata
\end{deluxetable}

In Figure~1 we plot the dispersion relation for the ${\rm f}^{\rm e}$
mode in the four models.  This is compared with the dispersion
relation for a purely polytropic disk and also with the limiting form
given by equation (\ref{eq12}).  It can be seen that the purely
polytropic dispersion relation provides a good approximation until the
wave action of the mode migrates into the atmosphere, at which point
the approximation given by equation (\ref{eq12}) becomes good.  The
value of $kH_s$ at which this transition occurs increases with
increasing $\tau$.  Note that the eigenvalue $\hat a$ is a negative
number whose magnitude increases with increasing $\tau$.  According to
the correspondence $\hat a=-n-{\textstyle{{1}\over{2}}}$, this means
that, in an optically thick disk, the atmospheric part of the
eigenfunction of the ${\rm f}^{\rm e}$ mode does not resemble the 2D
mode of a purely isothermal disk, but is more like the tail of a
high-order p mode after its final node.

\placefigure{fig1}

We now consider the propagation of the ${\rm f}^{\rm e}$ mode away
from an LR, as in Paper~I.  Away from the resonance, the radial flux
of angular momentum associated with the wave (averaged over $t$,
integrated over $\phi$ and $z$),
\begin{equation}
F^{\rm(a)}={{\pi
rm}\over{k}}\left({{\hat\omega^2-\kappa^2}\over{\hat\omega^2}}\right)
\int\rho|u|^2\,dz,
\end{equation}
is independent of $r$, and this condition determines the relative
normalization of the wave at each radius.  Our method of estimating
the nonlinearity of the wave at each radius is to locate the value of
$z$ at which the peak of the density of angular momentum wave action
(averaged over $t$),
\begin{equation}
{\cal
A}^{\rm(a)}=\left({{m}\over{2\hat\omega}}\right)\rho\left(|u|^2+|w|^2\right),
\end{equation}
occurs, and to compute the (isothermal) Mach number of the RMS velocity
perturbation,
\begin{equation}
{\cal M}=\left[{{{\textstyle{{1}\over{2}}}\left(|u|^2+|v|^2+|w|^2\right)}
\over{p/\rho}}\right]^{1/2},
\end{equation}
at that height.

Since this measure depends on the overall amplitude of the wave, we
normalize it in terms of its value at the resonance as follows.  In
the neighborhood of the LR, the horizontal components of the velocity
dominate, and the peak of the wave action density occurs on $z=0$.
The form of the inner solution is (cf. Paper~I)
\begin{eqnarray}
u&\sim&C\tilde u(z)\left[Ai(qx)\pm iGi(qx)\right],\\
v&\sim&iC\tilde v(z)\left[Ai(qx)\pm iGi(qx)\right],
\end{eqnarray}
where $x=(r-r_{\rm L})/r_{\rm L}$, $C$ is a constant, $q$ and $\tilde
u(z)$ are the f-mode eigenvalue and eigenfunction discussed in
Paper~I, and $\tilde v=-(2B/\hat\omega)\tilde u$.  The constant $C$ is
determined by asymptotic matching to the WKB solution away from the
LR.  Now the maximum value of $[Ai(qx)]^2+[Gi(qx)]^2$ is approximately
$0.3603$ and occurs at $qx\approx-1.845$.  This determines the maximum
Mach number at the LR, which we call ${\cal M}_{\rm L}$.  The scalings
are such that ${\cal M}/{\cal M}_{\rm L}$ away from resonance is
$O\left((H_s/r)^{1/6}\right)$, and for this reason we evaluate the
quantity
\begin{equation}
\left({{{\cal M}}\over{{\cal M}_{\rm
L}}}\right)\left({{r}\over{H_s}}\right)^{1/6}.\label{eq14}
\end{equation}

As in Paper~I, we have assumed that the disk is Keplerian, so that the
dimensionless intrinsic frequency of the wave is given by
\begin{equation}
{{\hat\omega}\over{\Omega}}=m\left[\left({{r}\over{r_{\rm c}}}\right)^{3/2}
-1\right],
\end{equation}
where $r_{\rm c}$ is the corotation radius of the mode.  We have also
assumed that $H_s\propto H\propto r$ and $\Sigma\propto r^{-1}$.  This
variation of parameters was selected so that the wave Mach number
would be finite at the radial center of the disk.  In this way, we
have removed geometrical focusing effects from the wave-channeling
effects of interest in this paper.

In Figure~2 we plot the variation with $r$ of the height $z_{\rm
peak}$ of the local vertical peak in the wave action density.  We
consider both inward propagation from the inner LR and outward
propagation from the outer LR; we consider azimuthal wave numbers
$m=2$ and $m=10$; and we consider the four models with different
optical depths.  In Figure~3 we plot the quantity in equation
(\ref{eq14}) for each of these cases.  As Figure~2 shows, the f mode
is launched at resonance with the peak of wave action density located
at the mid-plane and the peak rises away from the mid-plane as it
propagates away from the resonance.  When $z_{\rm peak}$ reaches the
base of the atmosphere, it resides there over a non-zero interval in
radius, as is apparent in Figure~2 by the flat slope of the curves.
This discontinuous behavior of the slope is caused by our idealization
of the transition from the polytropic interior to the isothermal
atmosphere as occurring abruptly in the vertical direction.  Notice
that the rise of $z_{\rm peak}$ with distance from the resonance
increases with optical depth.  However, the peak never rises to great
heights in the atmosphere.  Figure~4 shows how the wave action density
is distributed vertically.

\placefigure{fig2}

\placefigure{fig3}

\placefigure{fig4}

We have verified that the waves are not artificially confined within
the disk in our model, either by the discontinuity in the sound speed
at the interface (which results from a change in the effective
adiabatic exponent) or by the discontinuity in the buoyancy frequency.
If the isothermal layer is treated as having an adiabatic exponent
equal to that of the polytropic layer, so that the sound speed is
everywhere continuous, the wave migrates into the atmosphere at nearly
the same stage during the radial propagation.  The peak of the wave
action density rises only slightly higher than in the case of a truly
isothermal atmosphere.  If, instead, the disk is treated as
adiabatically stratified in both layers, so that the buoyancy
frequency is everywhere continuous (and zero), the results are almost
indistinguishable from those we have presented here.  We emphasize
that radially propagating waves in an accretion disk are generically
confined in the vertical direction because the vertical gravity
increases with height above the mid-plane.

\section{Summary and discussion}

We have analyzed the effects of a disk atmosphere on the propagation
of a wave, the f mode, launched from a Lindblad resonance in a
thermally stratified disk. The atmosphere was modeled as an isothermal
layer that resides above a polytropic region that extends down to the
mid-plane.  The polytropic region represents the (vertically)
optically thick interior of the disk.  The effects of the atmosphere
can be important because the f mode becomes progressively more
confined to the disk surface with distance from resonance.  The wave
energy is concentrated in the vertical direction as the wave
propagates radially (see Figure 4). The peak of wave action density in
the disk follows the trajectories shown in Figure 2 for various values
of azimuthal wavenumber $m$ and disk optical depth $\tau$.  The
dominant amplification of a wave occurs in the thermally stratified
region below the disk atmosphere through the process of wave
channeling, as was anticipated in Paper I (see the discussion in
\S8.2). The disk atmosphere lessens the degree of wave amplification.

As was found in Paper I, the launched wave (f mode) behaves more like
a surface gravity wave as it propagates radially away from the
resonance.  In a purely vertically polytropic disk with no atmosphere,
the vertical extent of the surface gravity wave decreases with
increasing radial distance from resonance.  As a result, there is
usually a substantial increase in wave amplitude on a radial scale of
order $r_{\rm L}/m$ from resonance, through a process termed wave
channeling.  In the presence of an atmosphere, the effects of wave
channeling are lessened (see Figure 3).  The increase in wave
amplitude becomes limited once the wave enters the disk atmosphere.
For disks of smaller optical depths, the thermal stratification is
small and the wave enters the isothermal layer on a radial scale
somewhat smaller than $r_{\rm L}/m$.  Consequently, the wave behaves
more like the 2D acoustic mode at most radii, in accordance with the
expected limiting behavior of a vertically isothermal disk, in which
the wave is 2D at all radii and its peak never rises above the
mid-plane.

For disks around individual stars (circumstellar disks) in close
binary star systems ($m = 2$), the increase in wave Mach number from
resonance to disk center is typically less than a factor of 10, and
somewhat less for lower optical depths (see Figure 3a).  For disks
around binary star systems (circumbinary disks) ($m = 2$) the increase
in wave Mach number from resonance is large over a scale of order the
resonance radius (see Figure 3c).  For waves that are mildly nonlinear
at resonance, nonlinear wave damping is likely to be important in such
disks of modest vertical optical depth ($\tau > 10$).  These effects
are much stronger for circumbinary, rather than circumstellar disks.

For protostellar disks ($m \sim 10 $, see Figure 3b and 4b),
planet-forming regions are expected to be optically thick (Bell
et~al. 1997).  Typical wave velocities near a giant planet are likely
to be mildly supersonic, with Mach number around 0.2 (Goldreich \&
Tremaine 1979). In such a situation, wave channeling would likely lead
to shocks for disks of moderate optical depths (of order 100) on a
radial scale of a few times the disk thickness.

We emphasize that the detailed nonlinear outcome of the
wave-channeling process remains uncertain.  While the wave action is
concentrated in the polytropic layer, the motion is approximately
incompressible, but this is no longer true once the wave migrates into
the atmosphere, where shocks may limit the amplitude.  For this reason
we have used the Mach number of the wave motion as a measure of
nonlinearity.  The development of shocks or other nonlinear wave
phenomena under these circumstances could be usefully investigated
using numerical simulations.

\acknowledgments

We gratefully acknowledge support from NASA Origins of Solar Systems
Grants NAGW-4156 and NAG5-4310, and NATO travel grant CRG940189.  We
acknowledge the support and hospitality of the Isaac Newton Institute
for Mathematical Sciences.  We thank Jim Pringle for many discussions
on this problem.

\figcaption[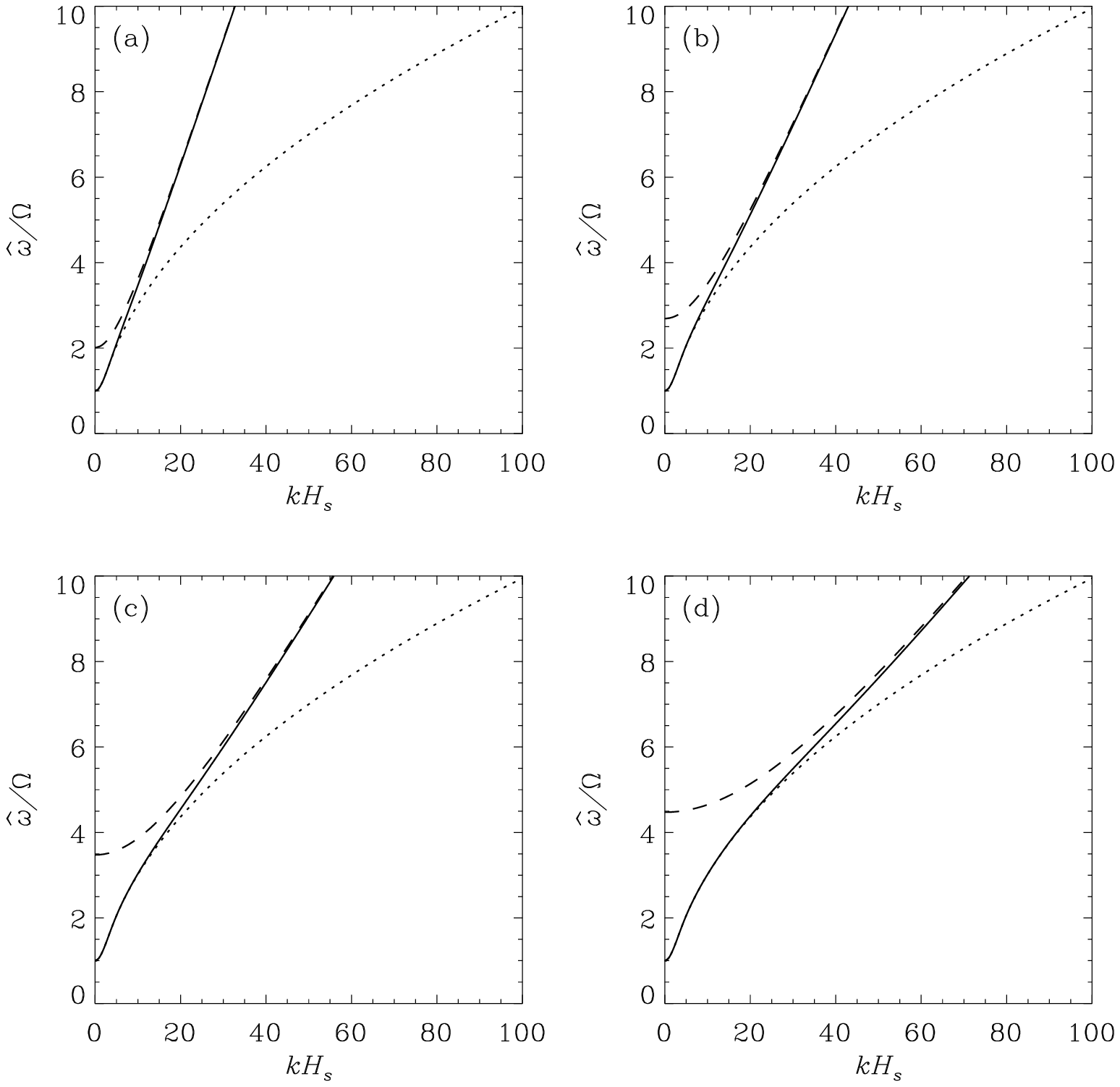]{Local dispersion relation ({\it solid line\/})
for the ${\rm f}^{\rm e}$ mode in a Keplerian disk consisting of a
polytropic layer with $s=3$ and $\gamma=5/3$ matched to an isothermal
atmosphere.  The four panels correspond to optical depths $\tau=10$,
100, 1000, and 10000.  {\it Dotted line\/}: dispersion relation for a
purely polytropic disk ($\tau=\infty$).  {\it Dashed line\/}:
asymptotic approximation given in equation (\ref{eq12}).\label{fig1}}

\figcaption[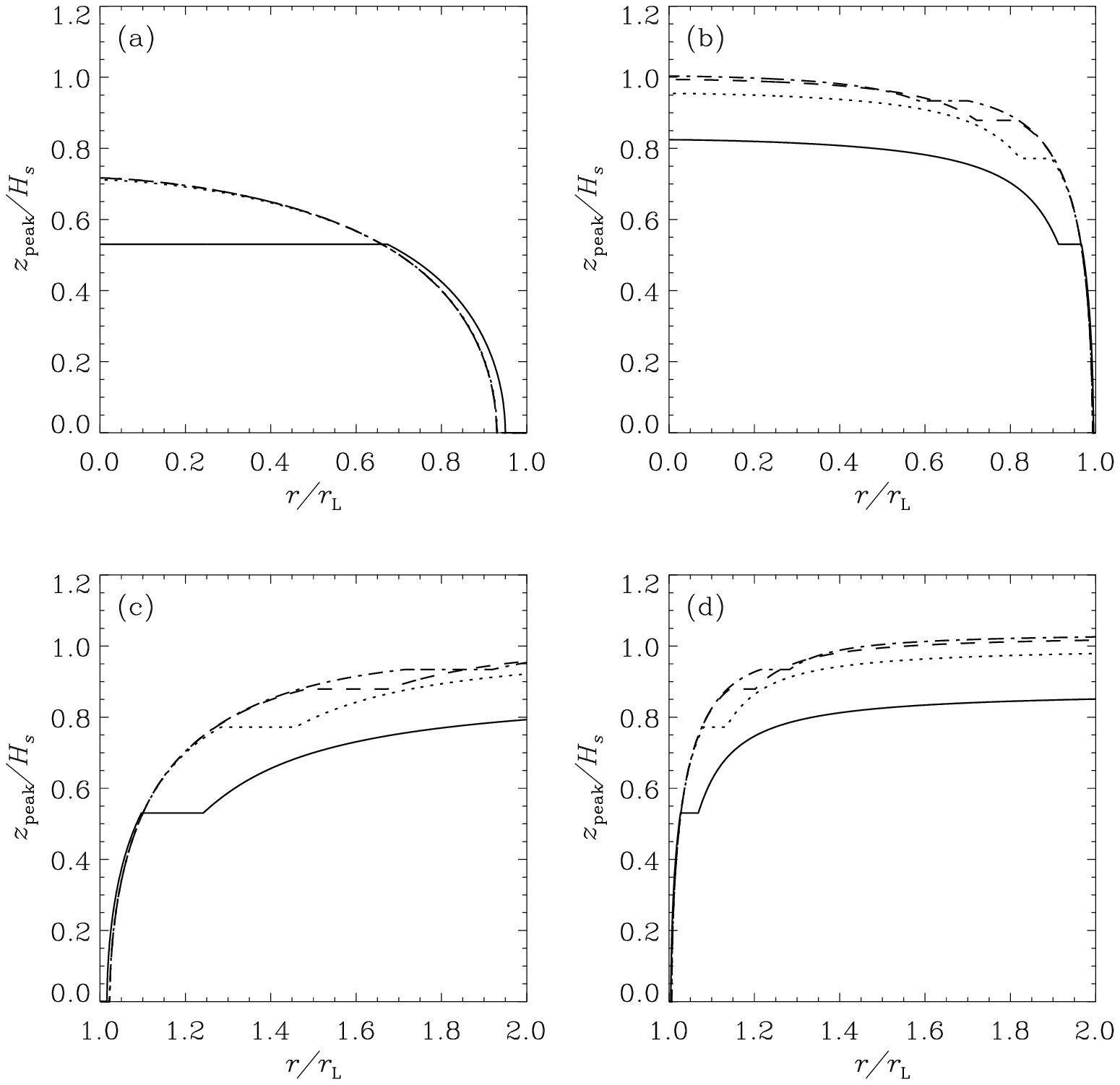]{Height of the peak of the wave action density as
the ${\rm f}^{\rm e}$ mode propagates away from an LR.  {\it Solid
line\/}: $\tau=10$.  {\it Dotted line\/}: $\tau=100$.  {\it Dashed
line\/}: $\tau=1000$.  {\it Dot-dashed line\/}: $\tau=10000$.  {\it
Panel (a)\/}: inner LR, $m=2$.  {\it Panel (b)\/}: inner LR, $m=10$.
{\it Panel (c)\/}: outer LR, $m=2$.  {\it Panel (d)\/}: outer LR,
$m=10$.\label{fig2}}

\figcaption[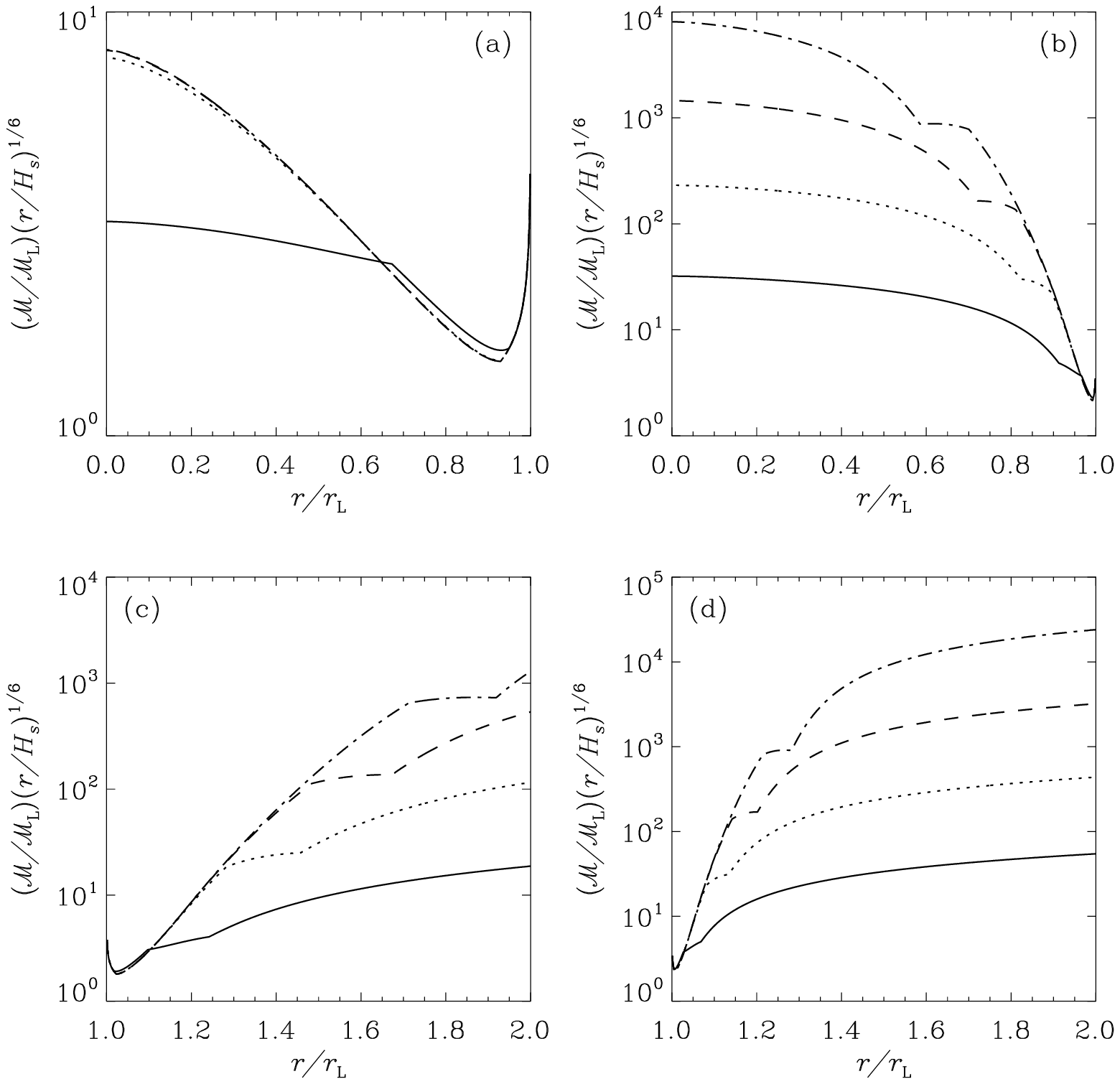]{RMS Mach number of the velocity perturbation at
the peak of the wave action density, relative to its value at the
resonance.  The vertical scale is logarithmic.  {\it Panel (a)\/}:
inner LR, $m=2$.  {\it Solid line\/}: $\tau=10$.  {\it Dotted line\/}:
$\tau=100$.  {\it Dashed line\/}: $\tau=1000$.  {\it Dot-dashed
line\/}: $\tau=10000$.  {\it Panel (b)\/}: inner LR, $m=10$.  {\it
Panel (c)\/}: outer LR, $m=2$.  {\it Panel (d)\/}: outer LR,
$m=10$.\label{fig3}}

\figcaption[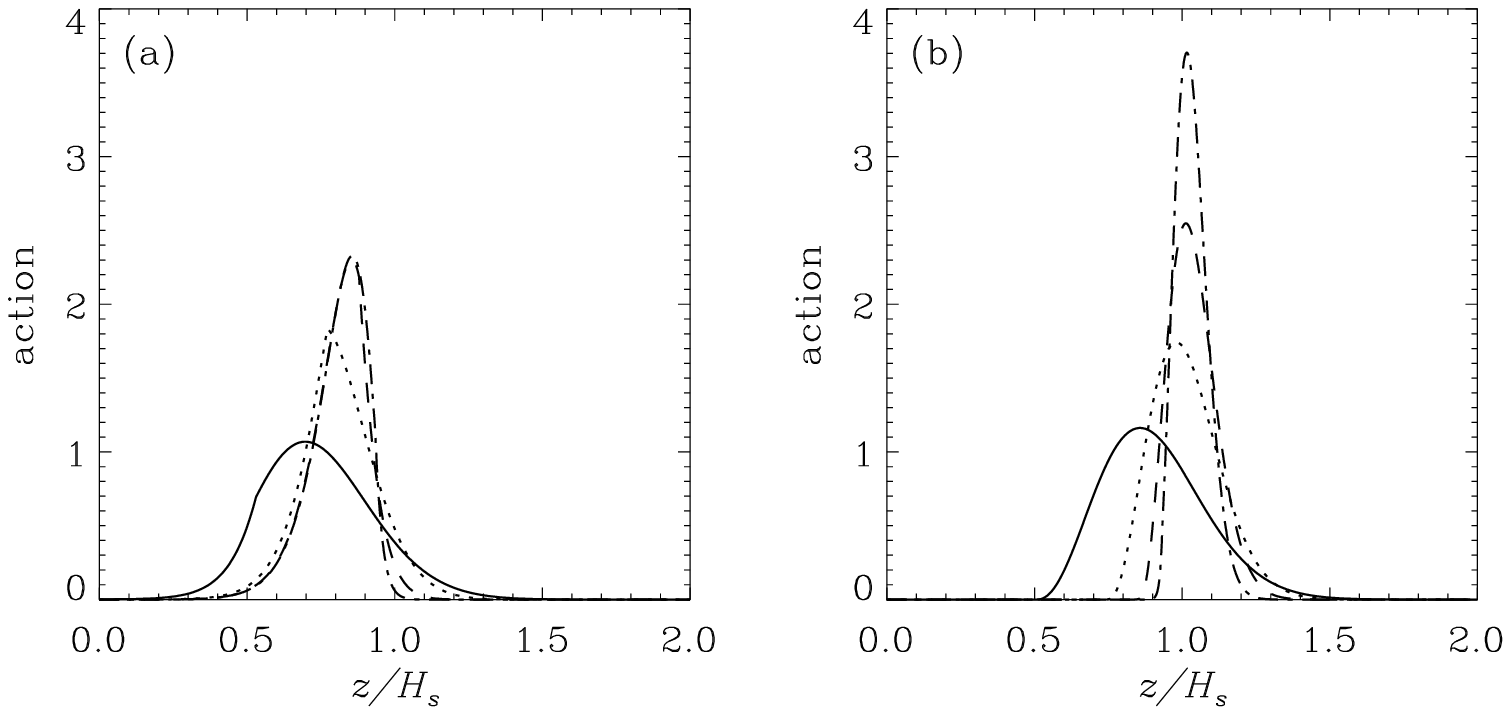]{Eigenfunctions of the ${\rm f}^{\rm e}$ mode.
The wave action density is plotted as a function of the vertical
coordinate for two different values of the radial wavenumber.  The
modes are normalized so that the wave action density integrates to
unity over the full vertical extent of the disk.  {\it Panel (a)\/}:
$kH_s=10$.  {\it Solid line\/}: $\tau=10$.  {\it Dotted line\/}:
$\tau=100$.  {\it Dashed line\/}: $\tau=1000$.  {\it Dot-dashed
line\/}: $\tau=10000$.  {\it Panel (b)\/}: $kH_s=100$.\label{fig4}}

\newpage
\pagestyle{empty}

\centerline{\epsfbox{fig1.eps}}

\vfill
\leftline{\bf Figure 1}

\newpage

\centerline{\epsfbox{fig2.eps}}

\vfill
\leftline{\bf Figure 2}

\newpage

\centerline{\epsfbox{fig3.eps}}

\vfill
\leftline{\bf Figure 3}

\newpage

\centerline{\epsfbox{fig4.eps}}

\vfill
\leftline{\bf Figure 4}

\end{document}